\documentclass[conference]{IEEEtran}
\IEEEoverridecommandlockouts
\usepackage{cite}
\usepackage{amsmath,amssymb,amsfonts}
\usepackage{algorithmic}
\usepackage{graphicx}
\usepackage{textcomp}
\usepackage{xcolor}
\usepackage{algorithm}
\usepackage{algorithmic}
\usepackage{hyperref}
\usepackage{float}
\DeclareMathOperator*{\argmax}{arg\,max}
\DeclareMathOperator*{\argmin}{arg\,min}
\begin{document}
\title{Multi-Rate Task-Oriented Communication for Multi-Edge Cooperative Inference
}

\author{
Dongwon~Kim, Jiwan~Seo, and Joonhyuk~Kang\\
School of Electrical Engineering, KAIST, Daejeon, South Korea\thanks{This work was partly supported by the Institute of Information \& Communications Technology Planning \& Evaluation (IITP)-ITRC (Information Technology Research Center) grant funded by the Korea government (MSIT) (IITP-2026-RS-2020-II201787, contribution rate: 50\%); in part by the Institute of Information \& Communications Technology Planning \& Evaluation (IITP) under 6G$\cdot$Cloud Research and Education Open Hub grant funded by the Korea government (MSIT) (IITP-2026-RS-2024-00428780, contribution rate: 50\%).}
} 
\maketitle
\begin{abstract}
   The integration of artificial intelligence (AI) with the Internet of Things (IoT) enables task-oriented communication for multi-edge cooperative inference system, where edge devices transmit extracted features of local sensory data to an edge server to perform AI-driven tasks. However, the privacy concerns and limited communication bandwidth pose fundamental challenges, since simultaneous transmission of extracted features with a single fixed compression ratio from all devices leads to severe inefficiency in communication resource utilization. To address this challenge, we propose a framework that dynamically adjusts the code rate in feature extraction based on its importance to the downstream inference task by adopting a rate-adaptive quantization (RAQ) scheme. Furthermore, to select the code rate for each edge device under limited bandwidth constraint, a dynamic programming (DP) approach is leveraged to allocate the code rate across discrete code rate options. Experiments on multi-view datasets demonstrate that the proposed frameworks significantly outperform the frameworks using fixed-rate quantization, achieving a favorable balance between communication efficiency and inference performance under limited bandwidth conditions. 
\end{abstract}

\section{Introduction}
The remarkable advancements in artificial intelligence (AI) have envisioned the ubiquitous deployment of intelligent applications in the sixth-generation (6G) communication networks \cite{huawei2022_6g}. In modern Internet of Things (IoT) environments, a large number of spatially distributed edge devices sense the physical object from diverse viewpoints. Such multi-view sensing provides complementary observations of the same object, enabling more accurate and robust performance in various applications than relying on a single device’s limited observation \cite{su2015multi}. However, conventional communication systems focus on ensuring bit-level reliability. Consequently, directly applying these systems to multi-edge sensing networks may result in the transmission of redundant information, thereby incurring significant communication overhead.

To bridge the gap between conventional communication systems and intelligent inference tasks, the concept of task-oriented communication has been recently introduced \cite{gunduz2022beyond, mostaani2022task}. Unlike conventional systems designed to recover bit-level transmission, task-oriented communication focuses on conveying semantic information that is relevant to the downstream inference task. In a multi-edge cooperative inference system, where multiple distributed devices cooperatively perform a shared task, this framework becomes particularly beneficial \cite{shao2022task, malka2024decentralized}. By allowing each device to transmit extracted task-relevant latent features instead of raw sensory data, the overall system can efficiently utilize limited bandwidth while preserving data privacy and maintaining high inference accuracy.

Recently, split inference has gained increasing attention as an effective strategy for facilitating task-oriented communication in multi-edge networks \cite{cai2024task, huang2024dynamic}. In this framework, a trained deep neural network is partitioned into a lightweight encoder deployed on the edge devices and a deeper decoder module operated on the edge server. The encoder extracts compact latent feature representations from the raw captured data and transmits them to the server for high-level inference, allowing resource-constrained devices to leverage large-scale AI models.

To realize efficient task-oriented communication, several studies have explored the vector-quantized variational autoencoder (VQ-VAE) \cite{van2017neural} as a backbone architecture \cite{chao2025task, nemati2023vq} for the encoder-decoder neural network. By jointly learning the encoder, decoder, and quantization modules in an end-to-end manner, a communication system equipped with VQ-VAE quantizes the latent features such that they are compatible with digital communication. Furthermore, along the line of multi-edge inference research, semantic communication frameworks with over-the-air computing have been proposed to improve the efficiency of multi-device information fusion and to reduce transmission redundancy \cite{yilmaz2022over, liu2024semantic}. In addition, recent studies have investigated information fusion strategies that exploit the superposition property of waveforms over the multiple-access channel to enhance the overall performance of inference tasks \cite{liu2023over}.

Despite recent advances in task-oriented communication, existing approaches still face fundamental limitations in practical deployment. Over-the-air computing-based methods, which transmit analog signals through the channel, require strict synchronization among devices and remain incompatible with conventional digital systems. In contrast, VQ-based methods employ a fixed code rate during latent feature extraction, enforcing a single compression ratio across all edge devices. This inflexible structure restricts the adaptive transmission rate across devices, resulting in inefficient utilization of communication resources. Therefore, a key challenge in task-oriented communication for multi-edge networks lies in developing an adaptive resource allocation strategy that dynamically adjusts the code rate according to task relevance and available bandwidth while maintaining the desired inference accuracy.

In this paper, we propose a multi-code rate task-oriented communication framework for a multi-edge cooperative inference system built upon the rate-adaptive quantization (RAQ)~\cite{seo2025raq} scheme. The proposed approach enables a single encoder on each edge device to adapt to varying bandwidth constraints by adjusting its code rate. Specifically, each device can adaptively extract the latent feature without additional training or parallel neural networks for different code rates. To maximize inference accuracy under bandwidth constraints, a dynamic programming (DP) algorithm is employed to allocate discrete code rates based on the entropy-derived reward of the captured sensory data.

The rest of this paper is organized as follows. In Section \ref{sec:system_model}, we introduce a multi-edge cooperative inference system that employs the encoding and decoding process with the RAQ scheme. Section \ref{sec:edge_multi-view_code_rate_selection} introduces the resource allocation strategy in terms of code rate. Section \ref{sec:experimental_results} presents the performance comparison between the proposed framework and other baselines. Finally, Section \ref{sec:conclusion} summarizes this paper.

\section{System Model}
\label{sec:system_model}
\begin{figure*}
    \centering
    \includegraphics[width=\textwidth]{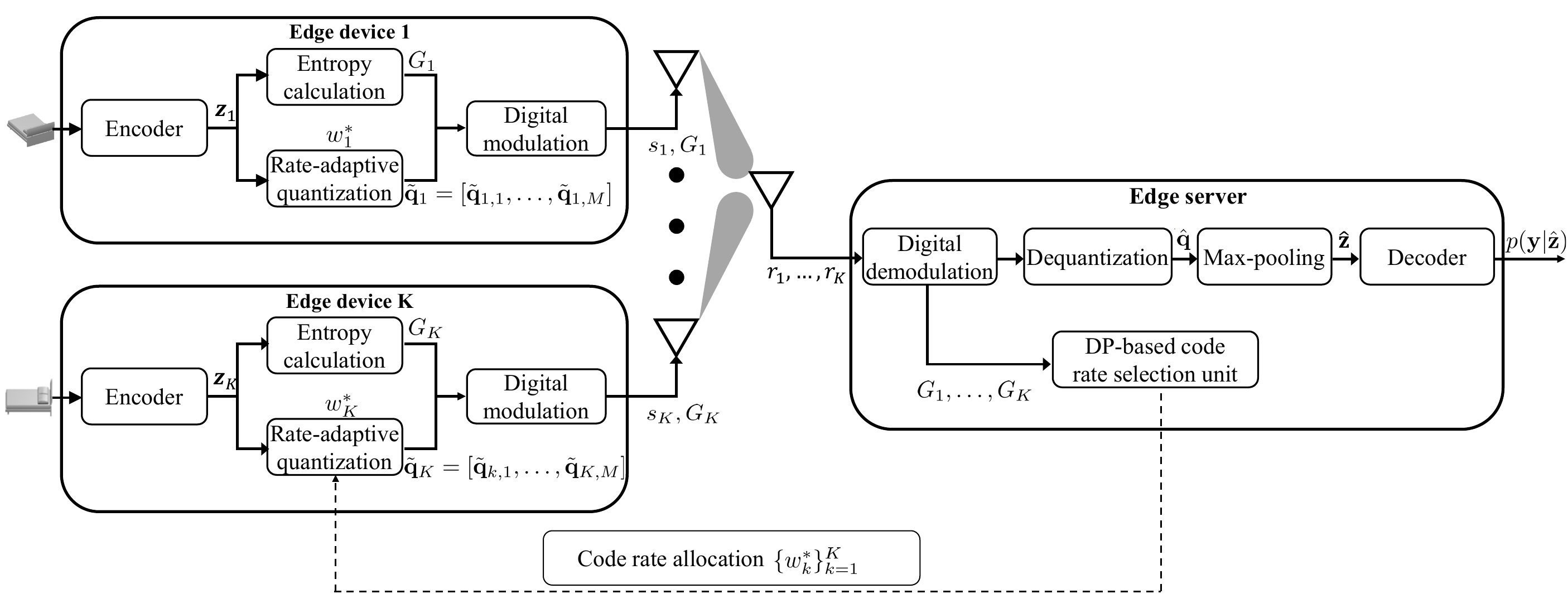}
    \caption{The architecture of the proposed multi-edge cooperative inference system. Each edge device encodes its captured view and computes the corresponding entropy value, while the edge server aggregates the encoded features and allocates the optimal code rate options $\{w^*_k\}_{k=1}^{K}$ for each edge device in the encoding process.}
    \label{fig:overall_outline}
\end{figure*}
In the proposed multi-edge cooperative inference system, as illustrated in Fig.~\ref{fig:overall_outline}, a total of $K$ edge devices and a single edge server collaboratively perform an inference task, such as image classification. Each edge device $k \in \mathcal{K} = {1,\ldots,K}$ captures a raw image $\mathbf{x}_k \in \mathbb{R}^{C\times H\times W}$ of a 3-D object from different viewpoints and orientation angles, where $C$, $H$, and $W$ denote the number of channels, height, and width of the image, respectively. The edge server then performs the final inference by aggregating the encoded feature transmitted from all edge devices. The objective of this cooperative framework is to accurately predict the true class probability vector $\mathbf{y}$ of the 3D object under limited communication bandwidth.

\subsection{Encoder at Edge Device}
The edge device is equipped with an encoder $f_{\text{enc}}(\mathbf{x}_k;\theta)$ parametrized by $\theta$, which extracts a latent feature $\mathbf{z}_k = f_{\text{enc}}(\mathbf{x}_k;\theta)\in \mathbb{R}^{d\times h\times w}$ from the raw input image $\mathbf{x}_k$. Here $d$, $h$, and $w$ are the feature dimension, height and width of latent feature $\mathbf{z}_{k}$, respectively. The latent feature vector $\mathbf{z}_k$ is factorized into $M$ sub-vectors $\{\mathbf{z}_{k,m}\}_{m=1}^{M}$. By applying vector quantization, each sub-vector of latent feature $\mathbf{z}_{k,m}$ is discretized through mapping it to the nearest codeword in a $\mathcal{E} = \{\mathbf{e}_j\}_{j=1}^{J}$ of size $J$, where each codeword $\mathbf{e}_j\in \mathbb{R}^{D}$ represents $D$-dimensional embedding vector. Accordingly, the quantized sub-vector of latent feature $\mathbf{q}_{k,m}$ is obtained as
\begin{align}
    \label{eq:quantization}
    \mathbf{q}_{k,m} = \argmin_i{||\mathbf{z}_{k,m}-\mathbf{e}_i||} \in \mathbb{R}^{D}.
\end{align}
For wireless transmission, each quantized sub-vector of latent feature $\mathbf{q}_{k,m}$ is represented by its corresponding codeword index, forming a vector of indices  $\mathbf{C} \in \{1,2,\ldots,J\}^{M}$. These indices are converted into the binary sequence and subsequently modulated into complex-valued symbols using standard digital modulation schemes (e.g., QPSK or 16-QAM), ensuring compatibility with conventional digital communication systems. We define the code rate as the number of bits required to represent each codeword index, which is given by $\log_2{J}$. Accordingly, transmitting all $M$ indices corresponding to the latent feature $\mathbf{z}_k$ for a single view requires a total of $M\log_2{J}$ bits.

The conventional VQ-VAE structure is inherently restricted by its reliance on a single fixed-size codebook, resulting in operation at a single code rate. In contrast, the proposed encoder equipped with the RAQ scheme enables flexible adjustment of the effective codebook size within a unified neural network architecture, thereby supporting multi-code rate in quantization of latent features according to the semantic importance of each view for the downstream inference task. The detailed implementation of the proposed framework with the RAQ scheme is illustrated in Section~\ref{sec:sec_raq}.

\subsection{Communication Model}
We consider orthogonal division multiple access (OFDMA) as the multiple access scheme to eliminate the interference among the multiple edge devices. In the proposed framework, the signals corresponding to each view are allocated to orthogonal subcarriers of the OFDMA system. The received signal on the $k$-th subcarrier $r_k$, corresponding to the $k$-th view, is expressed as   
\begin{align}
    r_k = h_ks_k + n_k,
\end{align}
where $s_k$ is the transmitted signal from the $k$-th edge server, $h_k$ is the flat-fading channel gain between the $k$-th edge device and edge server, and $n_k \sim \mathcal{CN}(0,\sigma_n^2)$ is the additive white Gaussian noise (AWGN). The channel gains $h_k$ are assumed to be known at the edge server, allowing channel equalization at the edge server. 

\subsection{Server-Side Decoding and Multi-View Feature Fusion}
In the multi-edge cooperative inference system, the objective of the edge server is to accurately perform the task based on the received information from the multiple edge devices, rather than recovering the individual message from multiple edge devices. To this end, the edge server employs the decoder that aims to predict the correct label of the observed 3-D object. 

The edge server first equalizes the received signal from the $k$-th edge device signal using the known channel gain obtained as  $\tilde{r}_k = r_k/h_k$. After equalization, the received signals are demodulated to obtain the binary sequence, which is than mapped to the corresponding codebook indices of all $M$ sub-vectors per view. From these indices, the recovered quantized sub-vectors of the latent features $\hat{\mathbf{q}}_{k,m}$ are retrieved from the shared codebook $\mathcal{E}$. The edge server  aggregates recovered sub-vectors from $K$ edge devices through element-wise max pooling, which retains the maximum value across views for each element in the $M$ sub-vectors. The aggregated sub-vector $\tilde{\mathbf{g}}_{m}$ is given by
\begin{align}
\tilde{\mathbf{g}}_{m}=\max{(\hat{\mathbf{q}}_{1,m}, \ldots, \hat{\mathbf{q}}_{k,m})},
\end{align}
where the maximum operation is applied element-wise over the recovered sub-vector of latent feature $\{{\hat{\mathbf{q}}_{k,m}}\}_{k=1}^{K}$ corresponding to the same sub-vector index $m$.

The recovered sub-vectors of latent feature $\{\tilde{\mathbf{g}}_m\}_{m=1}^{M}$ are concatenated to obtain the reconstructed latent feature $\hat{\mathbf{z}} = [\tilde{\mathbf{g}}_1, \tilde{\mathbf{g}}_2, \ldots, \tilde{\mathbf{g}}_M]$. The resulting reconstructed latent feature $\hat{\mathbf{z}}$ is then fed into the decoder $f_{\text{dec}}(\cdot;\phi)$ with learnable weight parameter $\phi$, which performs the final inference. The final inference results correspond to 
\begin{align}
    \hat{\mathbf{y}} = f_{\text{dec}}(\hat{\mathbf{z}};\phi),
\end{align}
where $\hat{\mathbf{y}}$ is the vector of predicted class probabilities. 

Max pooling is employed for multi-view feature aggregation, following the multi-view convolutional neural network (MVCNN) design \cite{su2015multi}, as it consistently outperforms average pooling approaches \cite{su2015multi, singhal2020communication}. Unlike average pooling, which averages feature activations across all views, max pooling retains the strongest activations, highlighting discriminative information while reducing redundancy. The aggregated latent feature $\hat{\mathbf{z}}$ thus constructs a view-invariant latent feature of the 3D object. 

\subsection{Rate-Adaptive Quantization (RAQ) Scheme for Multi-edge Inference System} 
\label{sec:sec_raq}
To enable multi-code rate control in latent feature extraction, we adopt a RAQ~\cite{seo2025raq} scheme, which allows dynamic adjustment of the code rate without modifying the encoder-decoder architecture. The core idea is to generate a variable-size codebook $\widetilde{\mathcal{E}}\!=\!\{\tilde{\mathbf{e}}_j\}_{j=1}^{\widetilde{J}}$ corresponding to the target code rate. To realize this, a sequence-to-sequence (Seq2Seq) model~\cite{sutskever2014sequence} sequentially generates the codewords of the codebook with target size $\widetilde{J}$, where the target size may be either smaller or larger than the original codebook size $J$. Specifically, RAQ leverages a long short-term memory (LSTM) network $f_{\text{LSTM}}(\cdot;\psi)$ parameterized by $\psi$ to produce codewords that compose the target codebook with target size $\widetilde{J}$. The overall generation process is expressed as 
\begin{equation}
p\big(\tilde{\mathcal{E}}|\mathcal{E},\widetilde{J}, \psi\big) = \prod_{j=1}^{\widetilde{J}} p\big(\tilde{\mathbf{e}}_j|\tilde{\mathbf{e}}_{<j}, \mathcal{E};\psi\big),
\label{eq:adaptive_codebook_gen}
\end{equation}
where $p(\tilde{\mathbf{e}}_j| \tilde{\mathbf{e}}_{<j},\mathcal{E};\psi)$ represents the conditional probability of generating the $j$-th codeword given the preceding ones and the base codebook $\mathcal{E}$. The sequential dependencies among codewords allow the model to generate a variable-size codebook that adapts to the target code rate. Furthermore, the resulting quantized latent feature for the $k$-th view is represented as $\tilde{\mathbf{q}}_k= [\tilde{\mathbf{q}}_{k,1},\ldots,\tilde{\mathbf{q}}_{k,M}]$, where each sub-vector $\tilde{\mathbf{q}}_{k,m}$ is obtained as in \eqref{eq:quantization} using the variable-size codebook $\tilde{\mathcal{E}}$ instead of the base codebook $\mathcal{E}$.

To achieve stable quantization and effective multi-view aggregation, the encoder-decoder pair and the Seq2Seq model are trained in two stages. In the first stage, single-view pretraining is conducted to establish a reliable quantization behavior before multi-edge aggregation. Specifically, the encoder and decoder networks are pretrained within a conventional VQ-based framework using single-view images only. During this phase, the encoder, base codebook, and decoder are jointly optimized using a fixed base codebook size $J$. This single-view pretraining process facilitates stable quantization and provides initial parameters of a model that serves as the pretrained model for subsequent multi-edge configuration.

In the second stage, the overall training incorporates the RAQ scheme into the multi-edge device setting, where all edge devices share a common encoder with shared parameters. During the forward pass, the quantized latent features from each device are transmitted to the decoder, where feature aggregation is performed via max-pooling. After obtaining the inference results from the decoder, the task-dependent loss is computed. During backpropagation, the loss gradients from all views are propagated through the shared encoder and accumulated to update its parameters, enabling view-invariant latent feature learning at the encoder while simultaneously improving inference accuracy at the decoder. To incorporate the RAQ scheme into the overall framework, the training objective is formulated as
\begin{equation}
\begin{split}
\mathcal{L}_{\text{RAQ}} = \mathcal{L}(\hat{\mathbf{y}},\mathbf{y}) \ +& \ \big|\big|\text{sg}\big[\mathbf{z}_k\big]-\tilde{\mathbf{q}}_k\big|\big|^2_2\!  \\
\ +& \ \big|\big|\mathbf{z}_k-\text{sg}\big[\tilde{\mathbf{q}}_k\big]\big|\big|^2_2,
\label{eq:raq_loss}
\end{split}    
\end{equation}
where $\mathcal{L}(\hat{\mathbf{y}}, \mathbf{y})$ denotes the task-dependent loss, such as the cross-entropy loss between the predicted and true class probabilities. The second and third terms represent the codebook and commitment losses, respectively, which jointly ensure consistent mapping between the encoder outputs and the LSTM-generated codewords. During training, the size of the generated codebook is randomly varied, while the codebook size corresponding to the target code rate is used during inference.
In practice, the overall training is performed with the conventional VQ loss, defined as $\mathcal{L}_{\text{VQ}} = \mathcal{L}(\hat{\mathbf{y}},\mathbf{y}) + ||\text{sg}[\mathbf{z}_k]\!-\!\mathbf{q}_k||^2_2 + ||\mathbf{z}_k\!-\!\text{sg}[\mathbf{q}_k]||^2_2$, so that both the base and variable-size codebook are trained in an end-to-end manner. This training process enables the model to dynamically construct a variable-size codebook that adapts to the target code rate, thereby achieving flexible feature compression without altering the underlying encoder-decoder architecture. The comprehensive realization of the RAQ scheme is presented in \cite{seo2025raq}.

\section{Code Rate Selection in Multi-Edge Cooperative Inference System}
\label{sec:edge_multi-view_code_rate_selection}
In this section, we illustrate the code rate selection strategy in the proposed framework, which optimizes the code rate of each view under the limited available bandwidth constraint. A dynamic programming (DP)-based approach is leveraged to choose the optimal code rate based on the entropy of the raw image. The entropy serves as the amount of information of the image, allowing the framework to  allocate different code rates to views based on their entropy level.  

\subsection{Entropy based Reward Estimation} 
To quantify the information content of each view, we compute the entropy based on the joint distribution of local grayscale features. Specifically, for every sliding window across the image, we form a pair $(l,m)$, where $l$ denotes the gray value of the center pixel and $m$ denotes the average gray value of all pixels within the sliding window. The occurrence frequency of each pair $f(l,m)$ is accumulated as the sliding window traverses the entire image. By normalizing the occurrence frequency $f(l,m)$ with the total number of pixels $H\!\cdot\!W$, we calculate the joint probability distribution as
\begin{equation}
        \label{eq:prob_entropy}
    p_{l,m} = f(l,m)/(H\cdot W).
\end{equation}
The entropy of the $k$-th view is then given by
\begin{align}
    \label{eq:entropy}
    G_k = -\sum_{(l,m)}p_{l,m}\log{p_{l,m}},
\end{align}
where the summation is taken over all possible pairs of gray values 
$l$ and the corresponding averaged values $m$. This formulation extends the conventional gray-level entropy calculation by incorporating local contextual information through the average value $m$, providing a more comprehensive measure of texture and structural variability. Such a definition is particularly effective in quantifying the information content of grayscale images \cite{wang2022ovpt}.

After estimating the entropy of each view, the resulting value is regarded as a reward that reflects the importance of the corresponding view to the overall inference at the edge server. Edge devices transmit a set of entropy values to the edge server. Based on the set of collected entropy values from all devices, the edge server dynamically assigns the code rate of each view so as to optimize overall inference accuracy under the limited communication bandwidth. The details of code rate selection are presented in the next section \ref{sec:DP}.
\begin{algorithm}[t]
\caption{DP-based Code Rate Selection}
\label{alg:alg1}
\begin{algorithmic}[1]
\STATE \textbf{Input:}  Number of views $K$, bandwidth budget $B$, discrete code rate option set $\mathcal{W}$, bandwidths $b_{k,i}$, rewards $\kappa_{k,i}$
\STATE \textbf{Output:}  Optimal code rate assignment $\{w_{k}^*\}_{k=1}^K$
\STATE Initialize DP table: $F(0,b) \text{ and } F(k,0)\gets 0$ for all $b \le B$ and $1 \le k \le K$
\STATE Initialize predecessor table: $\text{prev}(k,b) \gets \emptyset$
\FOR{$k = 1$ to $K$}
    \FOR{$b = 0$ to $B$}
        \STATE $F(k,b) \gets F(k-1,b)$ \hfill // no allocation to view $k$
        \FOR{each option $w_{k,i} \in \mathcal{W}$}
            \IF{$b_{k,i} \le b$}
                \STATE temp $\gets F(k-1, b - b_{k,i}) + \kappa_{k,i}$
                \IF{temp $> F(k,b)$}
                    \STATE $F(k,b) \gets$ temp
                    \STATE $\text{prev}(k,b,w) \gets (k-1, b - b_{k,i}, w_{k,i})$
                \ENDIF
            \ENDIF
        \ENDFOR
    \ENDFOR
\ENDFOR
\STATE Optimal reward value: $\Gamma = \max_{b \le B} F(K,b)$
\STATE Backtrack from $\Gamma$ using $\text{prev}$ to recover optimal code rate set $\{w_{k}^\ast\}_{k=1}^K$
\RETURN $\{w_{k}^\ast\}_{k=1}^K$
\end{algorithmic}
\end{algorithm}
\subsection{Dynamic Programming based Code Rate Selection}
\label{sec:DP}
Based on the entropy values in \eqref{eq:entropy}, the edge server selects code rates for each view from a discrete code rate option set $\mathcal{W} = \{w_i\}_{i=1}^{I}$. Choosing a code rate option $w_{k,i}$ requires a bandwidth $b_{k,i}$ proportional to the level of $w_{k,i}$ and yields a reward $\kappa_{k,i}$ derived from the entropy value $G_k$ of the corresponding edge device. The reward $\kappa_{k,i}$ represents the usefulness of the encoded latent feature for the inference task, where a higher code rate generally preserves richer visual information and consequently yields higher rewards. To account for the relative contribution of different code rates $w_{k,1}, w_{k,2}, \ldots, w_{k,I}$, empirical scaling factors $[\beta_1, \beta_2, \ldots, \beta_{I}]$ are introduced, where $\beta_i$ corresponds to $w_{k,i}$. Accordingly, the reward is defined as $\kappa_{k,i}=\beta_{i}\cdot G_k$, which effectively indicates the relative impact of each code rate on the overall task performance.

The system is constrained by a total bandwidth budget $B$, which prevents allocating the maximum code rate to all views simultaneously. The objective is to determine the optimal code rate assignment for each view that maximizes the overall reward while satisfying the budget constraint.

To this end, we formulate the optimization problem as follows,
\begin{align}
\label{eq:DP_formaulation}
\argmax_{\{w_{k,i} \in \mathcal{W}\}_{i=1}^K}  \sum_{i=1}^K r_{k,i}
\quad \text{s.t.} \quad
\sum_{i=1}^K b_{k,i} \le B.
\end{align}
We solve this problem using a DP approach. Specifically, we construct a DP table $F(k,b)$ of size $K \times B$, where each entry represents the maximum achievable reward by allocating bandwidth budget $b$ to the first $k$ views. The recursion is given by
\begin{align}
\label{DP_recursion}
F(k, b) = \max_{w_{k,i} \in \mathcal{W}}
\left\{ F(k-1, b - b_{k,i})\! +\! \kappa_{k,i} \big|  b_{k,i} \le b \right\},
\end{align}
with initialization $F(0,b)=0  \text{ and } F(k,0)=0$ for all $b \le B$ and $1 \leq k\leq K$. During each update of the DP table, the predecessor state is stored to reconstruct the optimal selection. Once the dynamic programming (DP) table is filled, the optimal reward is obtained as $\max_{b \le B} F(K, b)$, and the corresponding bandwidth assignment for each view is determined by backtracking the stored predecessors according to the bandwidth allocation $b_{k,i}$. Since each code rate option $w_{k,i}$ requires a corresponding bandwidth size $b_{k,i}$, the optimal code rate allocation for each view $\{w^*_k\}_{k=1}^{K}$ is accordingly selected based on the obtained bandwidth allocation. The allocated code rate for each view is fed back to the respective edge device, which encodes its captured view $\mathbf{x}_k$ using the assigned code rate. The overall procedure is summarized in Algorithm~\ref{alg:alg1}.


\section{Experimental Results}
\label{sec:experimental_results}
In this section, we evaluate the performance of the proposed framework in comparison with baselines. For the inference task, experiments are conducted on a classification task using the ModelNet10 dataset. The ModelNet10 dataset contains 10 object categories (e.g., chair, bed) with images of $224\times224$ resolution, each representing a 2D projection of a common 3D object captured from different viewpoints. During training, the MVCNN is trained with 6 views per object, whereas only up to 3 views are available during testing to emulate a limited observation scenario. The MVCNN architecture consists of an encoder with five convolutional layers and residual blocks, followed by a decoder with two fully connected layers. 
\begin{figure}[t]
    \centering
    \includegraphics[width=0.5\textwidth]{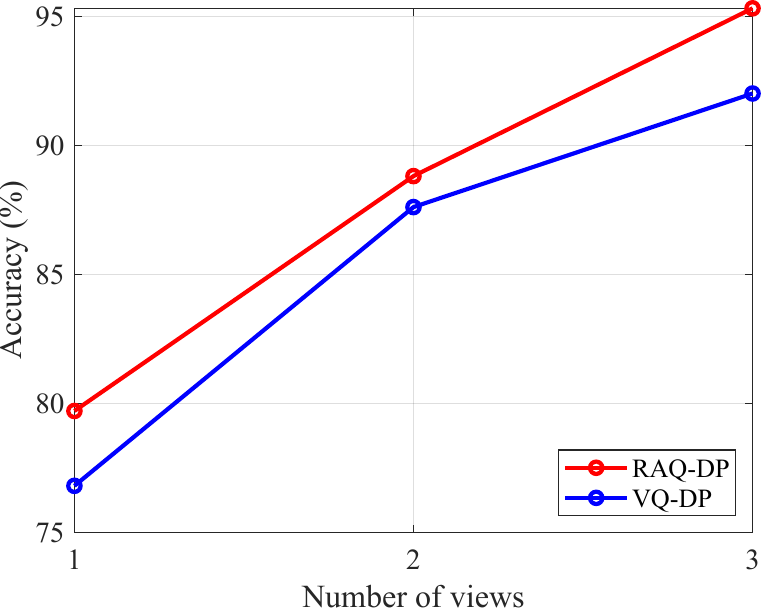} 
    \caption{Classification accuracy as a function of the number of views at 5dB SNR}
    \label{fig:per_view}
\end{figure}
\begin{figure}[t]
    \centering
    \includegraphics[width=0.5\textwidth]{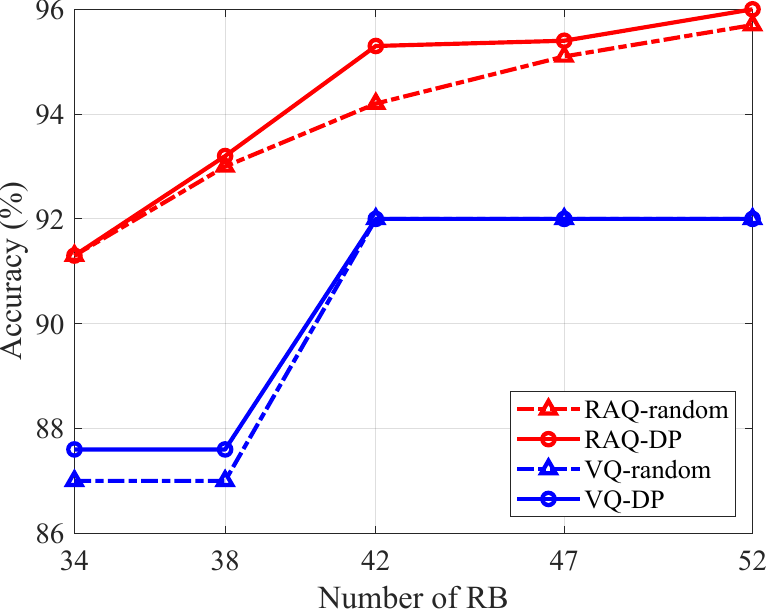} 
    \caption{Classification accuracy as a function of available bandwidth at 5dB SNR}
    \label{fig:per_rb}
\end{figure}
The communication protocol assumes that each edge device first transmits its estimated entropy $G_K$ to the edge server, which  determines the code rate and feeds it back to the devices. Each device then encodes its latent features using the assigned code rate. All experiments adopt QPSK digital modulation. The proposed RAQ framework supports three discrete code rate options corresponding to codebook sizes of $J=2^4,\,2^6,\,2^8$, forming a set size of $I = 3$. The scaling coefficients $\beta_i$ for each code rate are set to $0.75$, $0.85$, and $0.9$, respectively.

We define the resource allocation for each edge device in terms of the number of resource blocks (RBs). According to the 3GPP standard \cite{specification3GPP}, each RB spans 12 subcarriers over one subframe, consisting of two slots with seven OFDM symbols per slot, resulting in $168$ resource elements (REs). With QPSK modulation, each RE carries 2 bits, yielding 336 bits per RB. The total number of bits required to transmit the entire latent feature $\mathbf{z}_k$ is given by $M \times \log_2(J)$, where $M$ is set to $28 \times 28$ in all experiments. Accordingly, the required number of RBs for transmitting $\mathbf{z}_k$ is calculated as $\big\lceil {M \times \log_2(J)}/{336} \big\rceil$. For codebook sizes $J = 2^4$, $2^6$, and $2^8$, the corresponding RB requirements are 10, 14, and 19, respectively.

For comparison, we implement baseline models, which are variants of the proposed framework without the RAQ mechanism, adopting a VQ-based framework with a fixed code rate where the codebook size is set to $J = 2^6$. Furthermore, random code rate selection schemes are implemented for both RAQ- and VQ-based frameworks that fully utilize within the available bandwidth.

Fig.~\ref{fig:per_view} compares the classification accuracy of the RAQ- and VQ-based frameworks as a function of available views at a signal-to-noise ratio (SNR) of 5~dB. The channel gain is modeled as $h_k \sim \mathcal{CN}(0, \sigma_h^2)$ with a variance of $\sigma_h^2 = 1$, following a Rayleigh fading distribution. Under this setting, the SNR is defined as $||r_k||^2 / \sigma_n^2$. The available bandwidth is set to $14$, $28$, and $42$ RBs for one, two, and three views, respectively. As shown in Fig.~\ref{fig:per_view}, classification accuracy increases with the number of available views, as a larger number of views provides more complementary information. Moreover, the proposed RAQ-DP framework adaptively selects different code rates rather than using a fixed rate across all views, achieving higher accuracy than VQ-DP framework, as observed in the 3-view case.  

In Fig.~\ref{fig:per_rb}, we report the classification accuracy as a function of the number of available resource blocks (RBs) when SNR is set to 5 dB. The RAQ-based frameworks achieve higher accuracy as the number of RBs increases, while the VQ-based frameworks saturates beyond the 42 RBs. This performance gap arises because, within the RB range of 42 to 52, the VQ-based frameworks can only utilize a fixed code rate option corresponding to a codebook size of $2^6$ for every view, while the proposed RAQ-based framework flexibly adopts various code rate options. Notably, at RB sizes of 33 and 38, the proposed RAQ-based frameworks significantly outperforms both RAQ- and VQ-based frameworks when compared with their two-view results in Fig.~\ref{fig:per_view}, indicating that RAQ-based frameworks effectively utilize 3 views within this RB range, while VQ-based frameworks can only utilize 2 views. Furthermore, in the RAQ-based framework, the proposed DP-based code rate selection algorithm consistently outperforms random code selection across all RB sizes, demonstrating superior adaptability to bandwidth constraints through the entropy-based reward mechanism.

\section{Conclusion}
\label{sec:conclusion}
This paper has proposed a framework that supports multi-code rate task-oriented communication for multi-edge cooperative inference systems. The proposed framework enables each edge device to extract features with multiple code rates, allowing dynamic allocation of transmission resources based on the input data feature. A DP-based approach is employed to select the optimal code rate for each edge device according to captured data entropy under bandwidth constraints. Experimental results on multi-view datasets validate the effectiveness of the proposed RAQ framework with DP-based code rate selection, demonstrating superior performance over conventional VQ-based frameworks with fixed code rates. Future work will focus on extending this framework to dynamically adapt code rate allocation under varying channel conditions across different views.

\bibliographystyle{IEEEtran}
\bibliography{main}

\end{document}